\documentstyle[aaspp4,epsf]{article}
\def\pp{\par\parshape 2 0truecm 15.5truecm 1truecm 14.5truecm\noindent}
\newcommand{\simgt}{\lower.5ex\hbox{$\; \buildrel > \over \sim \;$}}
\newcommand{\simlt}{\lower.5ex\hbox{$\; \buildrel < \over \sim \;$}}
\begin{document}

\title{
Sunyaev -- Zel'dovich fluctuations 
from spatial correlations\\
between clusters of galaxies 
}

\bigskip

\author{
Eiichiro Komatsu
}

\affil{
Astronomical Institute, T\^{o}hoku University,\\
Aoba, Sendai 980-8578, Japan
}

\author{
Tetsu Kitayama
}

\affil{
Theoretical Astrophysics Group,
Department of Physics, Tokyo Metropolitan University,\\
Hachioji, Tokyo 192-0397, Japan
}

\bigskip

\begin{abstract}
  We present angular power spectra of the cosmic microwave background
  radiation anisotropy due to fluctuations of the Sunyaev-Zel'dovich
  (SZ) effect through clusters of galaxies. A contribution from the
  correlation among clusters is especially focused on, which has been
  neglected in the previous analyses.  Employing the evolving linear
  bias factor based on the Press-Schechter formalism, we find that the
  clustering contribution amounts to $20-30$\% of the Poissonian one
  at degree angular scales.  If we exclude clusters in the local
  universe, it even exceeds the Poissonian noise, and makes dominant
  contribution to the angular power spectrum.  As a concrete example,
  we demonstrate the subtraction of the {\em ROSAT} X-ray flux-limited
  cluster samples.  It indicates that we should include the clustering
  effect in the analysis of the SZ fluctuations.  We further find that
  the degree scale spectra essentially depend upon the normalization
  of the density fluctuations, i.e., $\sigma_8$, and the gas mass
  fraction of the cluster, rather than the density parameter of the
  universe and details of cluster evolution models.  Our results show
  that the SZ fluctuations at the degree scale will provide a possible
  measure of $\sigma_8$, while the arc-minute spectra a probe of the
  cluster evolution.  In addition, the clustering spectrum will give us
  valuable information on the bias at high redshift, if we
  can detect it by removing X-ray luminous clusters.
\end{abstract}

\section{Introduction}

The cosmic microwave background radiation (CMBR) anisotropy is now
known as the most powerful probe of our Universe (e.g. Hu, Sugiyama \&
Silk 1997).  In addition to the {\em primary} anisotropy which will
give us clues to most of the cosmological parameters within 10\%
accuracy (Bond, Efstathiou \& Tegmark 1997), there is the {\em
secondary} anisotropy caused by several processes in the low-redshift
universe, e.g., non-linear evolution of gravitational structures (Rees
\& Sciama 1968), peculiar velocity field of re-ionized clumps
(Ostriker \& Vishniac 1987), and the Sunyaev-Zel'dovich (SZ) effect
through clusters of galaxies.  The SZ effect is the brightness change
of the CMBR due to inverse Compton scattering by hot electrons
(Zel'dovich \& Sunyaev 1969; Sunyaev \& Zel'dovich 1972). So far the
brightness {\em decrement} in the Rayleigh-Jeans band has been
detected toward tens of clusters (see, Birkinshaw 1999 for review),
and we have recently detected the {\em increment} in the Wien band
(Komatsu et al. 1999).  Thanks to such specific
spectral features, one will be able to separate the SZ fluctuations
(SZF) rather accurately from any other signals using future satellite
observations, e.g., that of {\em PLANCK} surveyor (Hobson et al. 1998).
Furthermore, the SZF should provide a probe of high-$z$ clusters yet
to be detected otherwise, because the SZ brightness is free from the
redshift diminishing effect (Birkinshaw 1999) unlike the X-ray
brightness. 

Although several theoretical studies have been done on the SZF
(Rephaeli 1981; Ostriker \& Vishniac 1986; Cole \& Kaiser 1988;
Schaeffer \& Silk 1988; Markevitch et al. 1991; Makino \& Suto 1993;
Bartlett \& Silk 1994; Persi et al. 1995; Bond \& Myers 1996;
Atrio-Barandera \& M\"ucket 1999), they mainly focus on the Poissonian
component only and much less attention is paid to that arisen from the
cluster-cluster correlation. In this Letter, we point out that
the cluster-cluster correlation can in fact make a non-negligible
contribution to the SZF at degree scales and explicitly predict its
angular power spectra. We further investigate possibilities of
exploring the nature of density fluctuations, cluster gas
distributions, and clustering properties of high-redshift clusters, by
means of the SZF.

\section{Model}
\subsection{Angular Profile of a Cluster}

The gas density profile of each cluster is approximated by the
spherical isothermal $\beta$-model, $\rho_{\rm gas}(r)=\rho_{\rm gas0}
\left[1+\left(r/r_{\rm c}\right)^2\right]^{-3\beta/2}$, where $r_{\rm
c}$ is the core radius, and $\rho_{\rm gas0}$ is the central gas mass
density. To account for the fact that a real cluster has a finite
size, we introduce the Gaussian filter on the edge at radius $R$,
i.e., $\rho_{\rm gas}(r)\rightarrow \rho_{\rm gas}(r)e^{-r^2/\xi
R^2}$, where $\xi$ is a fudge factor to normalize the total gas mass
of a cluster. Assuming $\beta=2/3$ for simplicity, the gas mass
contained within $R$ is
\begin{equation}
  M_{\rm gas} \equiv \int_0^R dV \rho_{\rm gas}(r) =
  4\pi\rho_{\rm gas0} r_{\rm c}^3\left(p-\arctan p\right),
  \label{eq:massnorm1} 
\end{equation}
where $p\equiv R/r_{\rm c}$. On the other hand,  
\begin{equation}
  \int_0^{\infty} dV \rho_{\rm gas}(r) e^{-r^2/\xi R^2}
  = 4\pi\rho_{\rm gas0} r_{\rm c}^3 h(p),
  \label{eq:massnorm2}    
\end{equation}
where 
\begin{equation}
  h(p) \equiv p\sqrt{\frac{\pi}4\xi} -\frac{\pi}2e^{1/\xi p^2}{\rm
  Erfc}\left(\sqrt{1/\xi p^2}\right),
  \label{eq:hp}  
\end{equation}
and ${\rm Erfc}(u)$ is the complementary error function. Matching
equations (\ref{eq:massnorm1}) and (\ref{eq:massnorm2}) at $p \gg 1$ gives $\xi
= 4/\pi$.

The above density profile yields the angular temperature profile
projected on the sky due to the SZ effect in terms of the ``Compton
$y$-parameter'' (Zel'dovich \& Sunyaev 1969) as
\begin{eqnarray}
\frac{ {\mit\Delta}T(\theta)}{T_{\rm cmbr}} &=& g(x)y(\theta), \\
         y(\theta) &=&
\sigma_T n_{e0}r_{\rm c}\left(\frac{k_{\rm B} T_e}{m_ec^2}\right)   
 \times \frac{\pi e^{1/\xi p^2}}{\sqrt{1+\left(\theta/\theta_{\rm c}\right)^2}}
    {\rm Erfc}\left(
              \sqrt{\frac{1+(\theta/\theta_{\rm c})^2}
                         {\xi p^2}}
              \right), 
\label{eq:prof} 
\end{eqnarray}
where $T_{\rm cmbr}=2.728 {\rm K}$ (Fixsen et al. 1996), $g(x)\equiv
x\coth(x/2)-4$, $x\equiv h\nu/k_{\rm B}T_{\rm cmbr}$, $n_{e0}$ is the central
electron number density, $T_e$ is the electron temperature, and other
physical constants have their usual meanings. The angular core size 
$\theta_{\rm c}$ is related to $r_{\rm c}$ via $\theta_{\rm c} = r_{\rm c}/d_A(z)$, where
$d_A$ is the angular diameter distance to a cluster at redshift $z$.

We model that clusters have virialized at $z$ and relate the
quantities $R$, $T_e$, $n_{e0}$ and $r_{\rm c}$ to the total mass $M$ and
redshift $z$ based on the spherical infall model (Peebles 1980, \S
19). We assume that $R$ is equal to the virial radius,
$R=[3M/4\pi\rho_{\rm univ}(z)\Delta_{\rm c}(z)]^{1/3}$, where $\rho_{\rm
univ}(z)$ is the mean density of the universe, and $\Delta_{\rm c}(z)$ is
the mean density of a cluster in units of $\rho_{\rm univ}(z)$ given
in Lacey \& Cole (1993) and Nakamura \& Suto (1997) for an arbitrary
cosmology. Then the electron temperature is expressed as $k_{\rm B}T_e=\eta
\mu m_p GM/3R$, using the mean molecular weight $\mu$ and a fudge
factor of order unity $\eta$, for which we adopt $\mu=0.59$ and $\eta
= \beta^{-1} = 3/2$, respectively.
The central electron number density is derived with the help of
equation (\ref{eq:massnorm2}) as $n_{e0} = (\rho_{\rm gas0}/m_p)(X+1)/2 =
\left(M f_{\rm gas}/m_p4\pi r_{\rm c}^3 h(p)\right)(X+1)/2$,
where we fix the hydrogen abundance at $X=0.76$ and assume that the
gas mass fraction is equal to the cosmological mean, $f_{\rm
gas}=M_{\rm gas}/M = \Omega_{\rm b}/\Omega_0$ (White et al. 1993). Given
large uncertainties in the formation of cluster core regions, we
analyze two different scenarios for the evolution of $r_{\rm c}$, i.e., the
{\em self-similar} (SS) model (Kaiser 1986), and the {\em
entropy-driven} (ED) model (Kaiser 1991; Evrard \& Henry 1991; Bower
1997).  In the SS model, $r_{\rm c}$ is simply proportional to the virial
radius,
\begin{equation}
  \label{eq:pss}
  p_{\rm ss}(M,z) = {\rm constant}. 
\end{equation}
In the ED model, on the other hand, the core is in a minimum entropy
phase, $s_{\rm min} = c_V \ln(T_e/\rho_{\rm gas0}^{\gamma -1})$, and its
evolution is specified by $s_{\rm min}(z) = s_{\rm min}(z=0) + c_V
\epsilon\ln(1+z)$ with a free parameter $\epsilon$, yielding  
\begin{eqnarray}
  p_{\rm ed}(M,z)
  &\approx&
  \left(\frac{3\rho_{\rm gas0}(M,z)/f_{\rm gas}}
  {\rho_{\rm univ}(z) \Delta_{\rm c}(z)}\right)^{1/2} \nonumber \\
  &\propto& \frac{T_e^{\frac{1}{2(\gamma-1)}}(M,z)}
{\Delta_{\rm c}^{1/2}(z)}   (1+z)^{-\frac{3}{2}-\frac{\epsilon}{2(\gamma-1)}},
  \label{eq:ped}
\end{eqnarray}
where we have used $M =(4\pi/3)\rho_{\rm univ}(z) \Delta_{\rm c}(z)R^3
\approx 4\pi(\rho_{\rm gas0}/f_{\rm gas})r_{\rm c}^3 p$ for $p\gg 1$.
The adiabatic index is fixed at $\gamma=5/3$ throughout. Since
Mushotzky \& Scharf (1997) obtained the constraint of $\epsilon = 0\pm
0.9$ from X-ray cluster samples, we demonstrate three cases of
$\epsilon=-1,0,1$ in the same way as Atrio-Barandela \& M\"ucket
(1999).  We normalize the proportional constants in both models at
$M=10^{15}h^{-1}M_\odot$ and $z=0$ assuming $r_{\rm
c}(10^{15}h^{-1}M_\odot, z=0) = 0.15h^{-1}{\rm Mpc}$.

\subsection{Angular Power Spectrum}

One conventionally expands the angular two-point correlation function
of the temperature distribution in the sky into the Legendre
Polynominal:
\begin{equation}
  \label{eq:corr}
  \left<
  \frac{\mit{\Delta}T}{T_{\rm cmbr}}(\mbox{\boldmath$n$})
  \frac{\mit{\Delta}T}{T_{\rm cmbr}}(\mbox{\boldmath$n$}+\mbox{\boldmath$\theta$})
  \right>
  =\frac1{4\pi}\sum_l C_l P_l(\cos\theta).
\end{equation}
Since we are considering discrete sources, we can write
$C_l=C_l^{(P)}+C_l^{(C)}$, where $C_l^{(P)}$ is the contribution from the
Poissonian noise and $C_l^{(C)}$ is from the correlation among clusters
(Peebles 1980, \S 41).  For convenience, we define the power spectrum of
$y$-parameter, $C^{yy}_l \equiv C_l/g^2(x)$, which is independent of
frequency.  Following Cole \& Kaiser (1988) and employing Limber's equation in
the cosmological context (e.g., Peebles 1980, \S 56), we get an integral
expression of $C^{yy}_l{}^{(P,C)}$:
\begin{eqnarray}
  \label{eq:Cp}
  C^{yy}_l{}^{(P)} &=& \int_0^{z_{\rm dec}} dz \frac{dV}{dz}
                \int_{M_{\rm min}}^{M_{\rm max}} dM
                \frac{dn(M,z)}{dM} \left|y_l(M,z)\right|^2,\\
  \label{eq:Cc}
  C^{yy}_l{}^{(C)} &=& \int_0^{z_{\rm dec}} dz \frac{dV}{dz}
                       P_{\rm m}\left(k=\frac l{r(z)},z\right)
                       \left[\int_{M_{\rm min}}^{M_{\rm max}} dM
                       \Phi_l(M,z)\right]^2,
\end{eqnarray}
where $V(z)$ and $r(z)$ are the comoving volume and the comoving
distance, respectively, $\Phi_l(M,z)$ is the ``biased temperature
function'':
\begin{equation}
  \label{eq:phi}
  \Phi_l(M,z)\equiv \frac{dn(M,z)}{dM}b(M,z) y_l(M,z), 
\end{equation}
and $y_l$ is the angular Fourier transform of $y(\theta)$. In
equations (\ref{eq:Cp})
and (\ref{eq:Cc}), $z_{\rm dec}$ is the redshift of photon decoupling, and we
choose $M_{\rm min}=5\times 10^{13}h^{-1}M_\odot$ and $M_{\rm max}=5\times
10^{15}h^{-1}M_\odot$ to bracket the mass scale of clusters.  The mass
function $dn/dM$ is computed using the Press-Schechter formula (Press \&
Schechter 1974):
\begin{equation}
  \label{eq:PS}
  \frac{dn(M,z)}{dM}
  = \sqrt{\frac{\pi}2}\frac{\rho_{\rm univ}(0)}{M}
    \left|\frac{d\ln\sigma(M)}{dM}\right|
    \nu(M,z)
    e^{-\nu^2(M,z)/2}, 
\end{equation}
where $\sigma(M)$ is the present-day rms mass fluctuation within the top-hat
filter, $\nu(M,z)\equiv \delta_{\rm c}(z)/D(z)\sigma(M)$ is the ``peak-height
threshold'', $D(z)$ is the linear growth factor of density fluctuations, and
$\delta_{\rm c}(z)$ is the threshold overdensity of spherical collapse given
in Lacey \& Cole (1993) and Nakamura \& Suto (1997) for an arbitrary
cosmological model.  We assume that the matter power spectrum, $P_{\rm
m}(k,z)$, is related to the power spectrum of the cluster correlation
function, $P_{\rm c}(k,M_1,M_2,z)$, via the time-dependent linear bias factor,
$b(M,z)$, i.e., $P_{\rm c}(k,M_1,M_2,z)=b(M_1,z)b(M_2,z)D^2(z)P_{\rm
m}(k,z=0)$. We adopt for $b(M,z)$ an analytic formula derived by Mo \& White
(1996): $b(M,z) = 1 + \left(\nu^2(M,z)-1\right)/\delta_{\rm c}(z)$, which
shows an excellent agreement with numerical simulations at the cluster mass
scale (Jing 1998, 1999). To calculate $P_{\rm m}(k)$, we employ the
Harrison-Zel'dovich spectrum and the transfer function of Bardeen et
al. (1986) with the shape parameter given in Sugiyama (1995). Unless stated
explicitly, we assume the $\Lambda$CDM cosmology with $\Omega_0=0.3$,
$\lambda_0=0.7$, $\Omega_{\rm b}=0.05$, $h=0.7$ and $\sigma_8=1.0$, which
agrees with both the COBE normalization and the cluster abundance
(e.g. Kitayama \& Suto 1997).

\section{Results and Discussions}

\subsection{Poissonian and Clustering Contributions to the
Sunyaev-Zel'dovich Fluctuations}

\begin{figure}[ht]
\begin{center}
     \leavevmode\epsfxsize=14cm \epsfbox{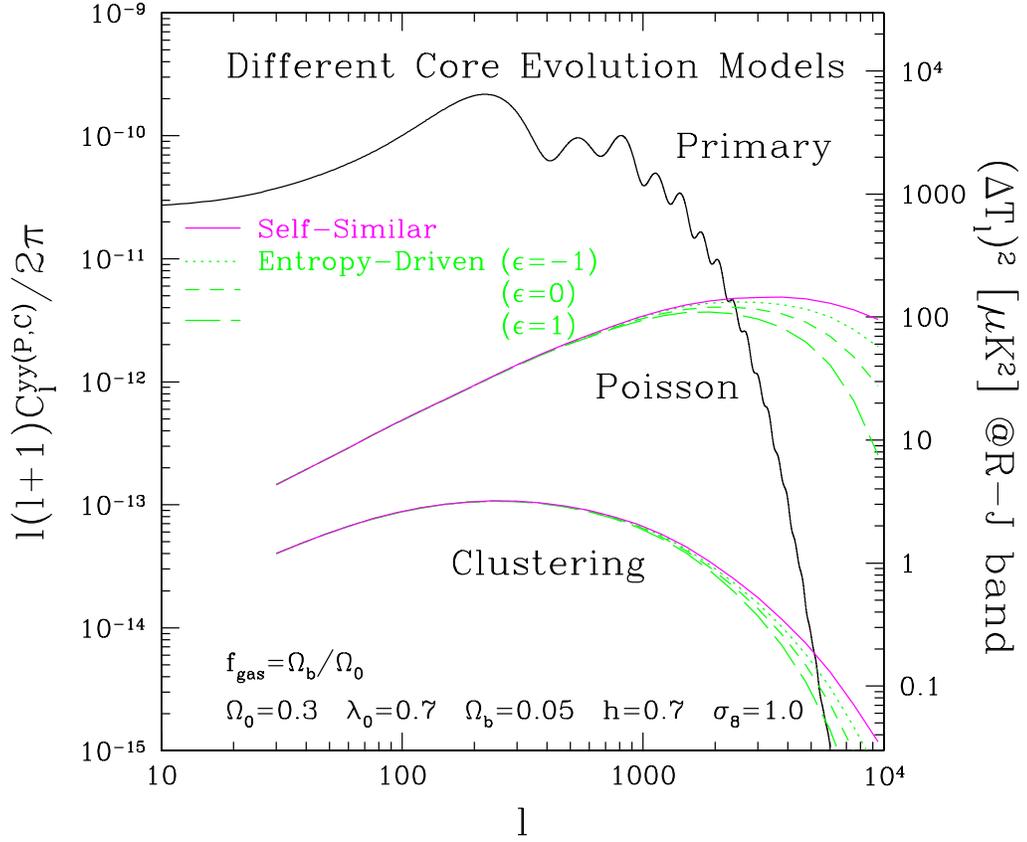}
\caption{
Angular power spectra of the Sunyaev-Zel'dovich fluctuations in terms
of $y$-parameter (see text for definitions). The Poissonian and
clustering spectra are shown for different core evolution
models. Solid lines represent the self-similar model, while others the
entropy-driven model with $\epsilon=-1,0,1$ from top to bottom. Also
plotted for reference is the primary temperature anisotropy 
expected in the Rayleigh-Jeans band ({\em right vertical axis}).
\label{fig:core}
}
\end{center}
\end{figure}

\begin{figure}[ht]
\begin{center}
    \leavevmode\epsfxsize=14cm \epsfbox{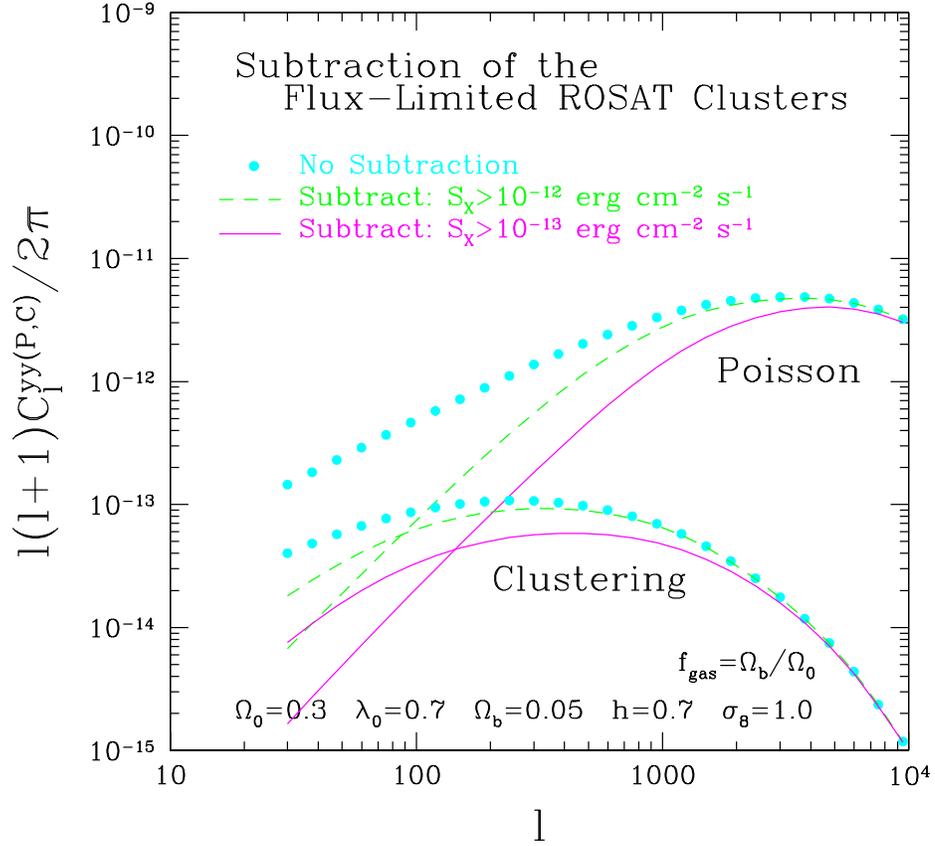}
\caption{
Effects of subtracting the clusters already resolved in the X-ray
band. Filled circles show the angular power spectra without the 
subtraction. Dashed and solid lines are the spectra after removing
the X-ray clusters above the flux limit of $10^{-12}$ and $10^{-13}\
{\rm erg\ cm^{-2}\ s^{-1}}$ in the $0.5-2$keV band, respectively.
\label{fig:rosat}
}
\end{center}
\end{figure}

Figure \ref{fig:core} shows the predicted power spectra of
$y$-parameter, $C^{yy}_l{}^{(P, C)}$, in the SS and three ED
($\epsilon=-1,0,1$) models.  The correlation term ${C}^{yy}_l{}^{(C)}$
amounts to around one-fifth of the Poissonian term
${C}^{yy}_l{}^{(P)}$ at $l\simlt 100$, and we will show later that
${C}^{yy}_l{}^{(C)}$ even dominates ${C}^{yy}_l{}^{(P)}$ in some
practical situations (Figure \ref{fig:rosat}).

For comparison, the primary anisotropy expected in units of the Rayleigh-Jeans
temperature is overlaid on the same figure. The SZF signals are shown to be
sub-dominant except at $l\simgt 2000$.  However, the simulations of
Hobson et al. (1998) indicate that the SZF can be separately
reconstructed from the CMBR map by the {\em PLANCK} satellite with
great accuracy especially at large angular scales ($l\simlt
500$). This is fully due to the specific spectral features of the SZ
effect in contrast to the primary signal and other secondary
sources. The angular resolution of the {\em PLANCK}, on the other
hand, seriously limits the accuracy of reconstruction at small angular
scales ($l\simgt 1000$).

The SZF spectra at $l\simlt 500$ is quite insensitive to the details of core
evolution models. This feature can be understood by looking at the angular
Fourier transformation of $y(\theta)$ (see eq. [\ref{eq:prof}]):
\begin{equation}
  \label{eq:proof}
  y_l\propto 
  \left\{
  \begin{array}{ll}
   n_{e0}r_{\rm c}\theta_{\rm c}e^{-l\theta_{\rm c}}/l
   & (l\theta_{\rm c}\gg 1),\\  
   n_{e0}r_{\rm c}\theta^2_{\rm c}
   & (l\theta_{\rm c}\ll 1).
  \end{array}
  \right.
\end{equation}
The spectrum at large angular scales is solely determined by the gas
mass: $y_{l\ll \theta_{\rm c}^{-1}} \propto M_{\rm gas}$, while the
small scale spectrum depends on the core radius: $y_{l\gg \theta_{\rm
    c}^{-1}} \propto M_{\rm gas}e^{-l\theta_{\rm c}}/(lr_{\rm c})$.
In what follows, therefore, we focus mainly on the large scale
spectrum and consider only the SS model.

In practical observations, one can remove from the reconstructed SZF
map bright clusters already detected in X-ray observations, e.g., by
{\em ROSAT} satellite.  Figure \ref{fig:rosat} illustrates the
expected SZF spectra after excluding the clusters above some threshold
X-ray fluxes, $S_X(0.5-2{\rm keV})=10^{-12}$ and $10^{-13}\ {\rm
  erg\ cm^{-2}\ s^{-1}}$. These values roughly correspond to the
limiting fluxes in the {\em ROSAT} All-Sky Survey (Ebeling et
al. 1997) and the {\em ROSAT} Deep Cluster Survey (Rosati et al.
1998). To compute the X-ray fluxes of clusters, we have used the same
method as described in Kitayama \& Suto (1997), assuming the following
form for the $L_X - T$ relation:
\begin{equation}
  \label{eq:LxT}
  L_X = L_{6\rm keV}\left(\frac{k_{\rm B}T}{6\ {\rm keV}}\right)^{\alpha}(1+z)^\zeta
  \ {\rm erg\ s^{-1}},
\end{equation}
with $L_{6\rm keV}=2.9\times 10^{44}h^{-2}$,
$\alpha=3.4$ and $\zeta=0$
(David et al. 1993; Mushotzky \& Scharf 1997),
and the X-ray band correction according to the emission model by Masai (1984).

We find that the Poissonian term ${C}^{yy}_l{}^{(P)}$ is greatly reduced by
the subtraction of X-ray bright clusters, and the correlation term
${C}^{yy}_l{}^{(C)}$ indeed dominates at $l<200$.  This reflects the greater
contribution of the correlation term ${C}^{yy}_l{}^{(C)}$ at higher $z$ and
smaller $l$, since ${C}^{yy}_l{}^{(C)}/{C}^{yy}_l{}^{(P)} \sim n(M,z)P_{\rm
c}(k=l/r(z)) \sim n(M,z)b^2(M,z)D^2(z)r(z)/l$, where $P_{\rm m}(k) \sim
k^{-1}$ on cluster scales. Therefore, we should take into account
${C}^{yy}_l{}^{(C)}$ in the analysis of the SZF, and it will possibly be
detected by the {\em PLANCK} with removing the {\em ROSAT} clusters.

Now, what does ${C}^{yy(C)}_{l\ll 1000}$ mean after this subtraction ?
It should represent the clustering properties of high-$z$ clusters
currently unresolved in X-ray observations.  It should be noted that
the above predictions depend on the linear bias assumption and the
evolution history of the bias factor.  Thus, if we could measure
${C}^{yy}_l{}^{(C)}$ separately from ${C}^{yy}_l{}^{(P)}$, it would in
turn provide the information on the bias at high-$z$.

\subsection{Implications on Cosmology}

\begin{figure}[ht]
\begin{center}
    \leavevmode\epsfxsize=14cm \epsfbox{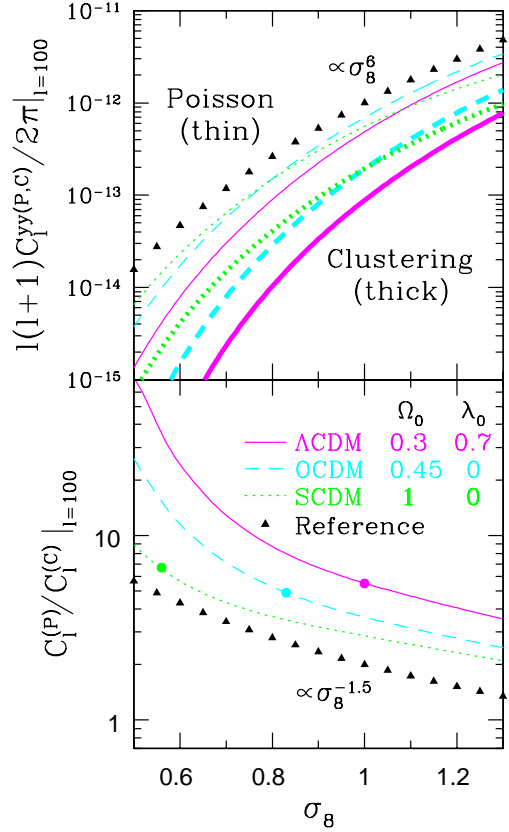}
\caption{
Dependences on $\sigma_8$ of angular power spectra at $l=100$ in three
representative cosmological models, $\Lambda$CDM ({\em solid lines}),
OCDM ({\em dashed}) and SCDM ({\em dotted}). The upper panel shows the
Poissonian ({\em thin lines}) and clustering ({\em thick lines}) power
spectra separately, while the lower panel illustrates ratios between
them.  Filled triangles indicate reference lines $\propto \sigma_8^6$
and $\sigma_8^{-1.5}$ in the upper and lower panels, respectively.
Filled circles in the lower panel represent the $\sigma_8$ value
inferred from the cluster abundance in each cosmological model
(Kitayama \& Suto 1997).
\label{fig:sigma8}
}
\end{center}
\end{figure}

Figure \ref{fig:sigma8} further illustrates the $\sigma_8$ dependence of
$C^{yy(P,C)}_{l=100}$ in three representative cosmological models:
$\Lambda$CDM, SCDM ($\Omega_0=1$, $\lambda=0$, $h=0.5$) and OCDM
($\Omega_0=0.45$, $\lambda=0$, $h=0.7$). The SZF spectra are extremely
sensitive to $\sigma_8$ while less sensitive to $\Omega_0$ and
$\lambda_0$. The correlation term $C_l^{(C)}$ is more sensitive to $\sigma_8$
than $C_l^{(P)}$, and can be fitted roughly by $C_l^{(C)}\propto
\sigma_8^{1.5} C_l^{(P)}$.  Since the {\em PLANCK} will provide the accurate
angular power spectrum of the SZF at the degree angular scales (Hobson et
al. 1998), they will provide a plausible measure of $\sigma_8$.  Furthermore,
figures \ref{fig:core} and \ref{fig:sigma8} suggest that the degree scale
spectrum is free from both unknown core physics of clusters and density
parameters of the universe.

\begin{figure}[ht]
\begin{center}
    \leavevmode\epsfxsize=14cm \epsfbox{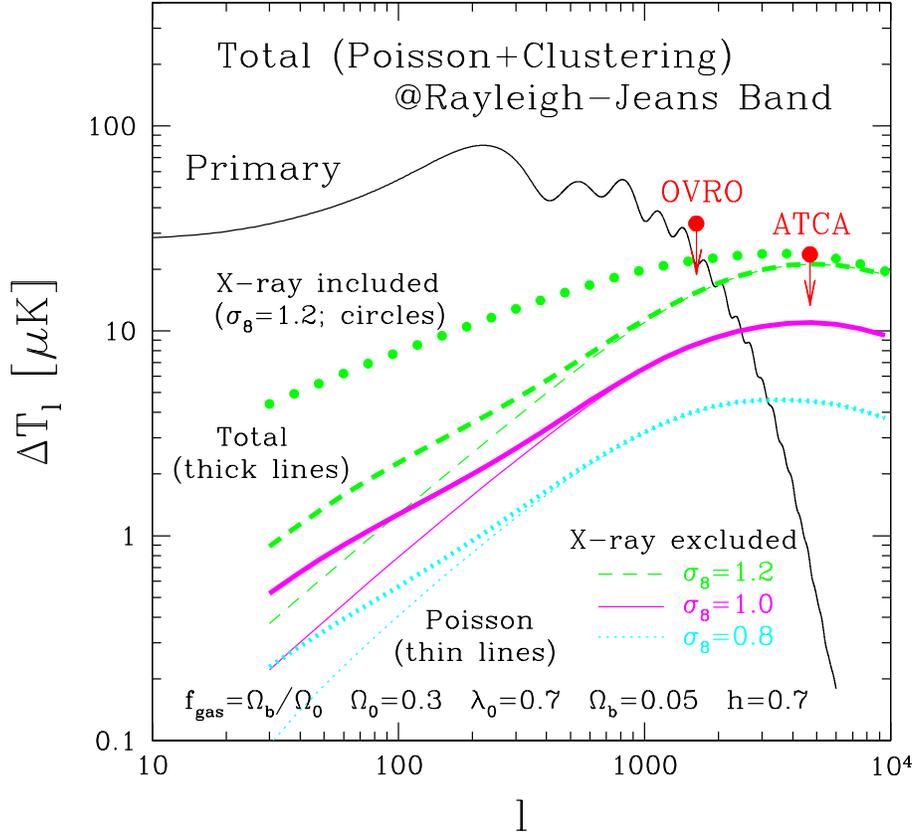}
\caption{
Total ({\em thick lines}) and Poissonian ({\em
thin lines}) angular power spectra of the Sunyaev-Zel'dovich fluctuations in
the Rayleigh-Jeans band excluding the X-ray bright clusters with $S_X(0.5-2
\mbox{keV}) >10^{-13}\ {\rm erg\ cm^{-2}\ s^{-1}}$.  Dotted, solid and dashed
lines represent the cases of $\sigma_8=0.8$, $1.0$ and $1.2$, respectively.
Circles show the total power spectrum with the X-ray bright clusters for
$\sigma_8=1.2$.  Also plotted for reference are the expected primary
anisotropy and the observed upper limits reported by OVRO (20 GHz) and ATCA
(8.7 GHz).
\label{fig:total}}
\end{center}
\end{figure}

Bond, Efstathiou \& Tegmark (1997) found that the expected $1\sigma$ error of
$\sigma_8$ determined by the {\em PLANCK} using the primary anisotropies alone
is as large as 20\%, which corresponds to changes in $C_l^{(P)}$ by a factor
of 3 and $C_l^{(C)}$ by a factor of 4.  Figure \ref{fig:total} shows the total
(Poissonian + clustering) and Poissonian angular power spectra of the SZF in
units of the Rayleigh-Jeans temperature, ${\mit\Delta T_l}\equiv T_{\rm
cmbr}\sqrt{l(l+1)C_l}$.  We show both cases of including and excluding the
X-ray bright clusters with $S_X>10^{-13}\ {\rm erg\ cm^{-2}\ s^{-1}}$.  It
illustrates how much the SZF change within 20\% error of $\sigma_8$.  Notice
that $C_l$ without X-ray clusters has a bending structure at $l\sim 200$,
below which $C^{(C)}_l$ exceeds $C^{(P)}_l$. The spectral shape also changes
with $\sigma_8$ because $C^{(P)}_l$ and $C^{(C)}_l$ depend differently on
$\sigma_8$. Such features should be quite helpful in detecting the clustering
contribution separately from the Poissonian one and probing the structure
formation in the high-$z$ universe.  For comparison, we also plot the primary
anisotropy spectrum and observational upper bounds obtained by OVRO (Readhead
et al. 1989) and ATCA (Subrahmanyan et al. 1998).

The SZF should play a complementary role to the primary anisotropies in the
determination of $\sigma_8$, and the clustering contribution hopefully brings
us fruitful information of high-$z$ bias.  It should be noticed,
however, that these results are based upon the assumption of $f_{\rm
gas}=\Omega_{\rm b}/\Omega_0$.  We found a rough scaling relation:
$C_l^{(P)}\propto f^2_{\rm gas}\sigma_8^6\Omega_0^2h$, at degree angular
scale, and it gives $C_l^{(P)}\propto \Omega_0^2$ when we specify $f_{\rm
gas}$ without cosmological parameters.  Uncertainties in $f_{\rm gas}$ yield
systematic errors in the amplitude of predicted SZF angular spectrum. These
points will be discussed in greater detail in future publications.


We thank Naoshi Sugiyama and Yasushi Suto for valuable suggestions,
Toshifumi Futamase, Makoto Hattori, Masahiro Takada and Keiichi Umetsu
for frequent discussions and critical comments.  We acknowledge
fellowships from Japan Society for the Promotion of Science.

\section*{REFERENCES}

\def\apjpap#1;#2;#3;#4; {\pp#1, {#2}, {#3}, #4}
\def\apjbook#1;#2;#3;#4; {\pp#1, {#2} (#3: #4)}
\def\apjppt#1;#2; {\pp#1, #2}
\def\apjproc#1;#2;#3;#4;#5;#6; {\pp#1, {#2} #3, (#4: #5), #6}

\apjpap Atrio-Barandela, F., \& M\"ucket, J. P. 1999;ApJ;515;465;
\apjpap Bardeen, J. M., Bond, J. R., Kaiser, N., \& Szalay,
A. S. 1986;ApJ;304;15;
\apjpap Bartlett, J. G., \& Silk, J. 1994;ApJ;423;12;
\apjpap Birkinshaw, M. 1999;Phys. Rep.;310;97;
\apjpap Bond, J. R., Efstathiou, G., \& Tegmark, M. 1997;MNRAS;291;L33;
\apjpap Bond, J. R., \& Myers, S. T. 1996;ApJS;103;1;
\apjpap Bower, R. G. 1997;MNRAS;288;355;
\apjpap Catelan, P., Lucchin, F., Matarrese, S., \& Porciani, C.
1998;MNRAS;297;692; 
\apjpap Cole, S., \& Kaiser, N. 1988;MNRAS;233;637;
\apjpap David, L. P., Slyz, A., Jones, C., Forman, W., Vrtilek,
S. D., \& Arnaud, K. A. 1993;ApJ;412;479; 
\apjpap Ebeling, H., Edge, A. C., Fabian, A. C., Allen, S. W., \&
Crawford C. S. 1997;ApJ;479;L101;
\apjpap Evrard, A. E., \& Henry, J. P. 1991;ApJ;383;95;
\apjpap Fixsen, D. J., Cheng, E. S., Gales, J. M. Mather, J. C.,
Shafer, R. A., \& Wright, E. L. 1996;ApJ;473;576;
\apjpap Hobson, M. P., Jones, A. W., Lasenby, A. N., \& Bouchet, F. R.
1998;MNRAS;300;1;
\apjpap Hu, W., Sugiyama, N., \& Silk, J. 1997;Nature;386;37;
\apjpap Jing, Y. P. 1998;ApJ;503;L9;
\apjpap Jing, Y. P. 1999;ApJ;515;L1;
\apjpap Kaiser, N. 1986;MNRAS;222;323;
\apjpap Kaiser, N. 1991;ApJ;383;104;
\apjpap Kitayama, T., \& Suto, Y. 1997;ApJ;490;557;
\apjpap Komatsu, E., Kitayama, T., Suto, Y., Hattori, M., Kawabe, R.,
Matsuo, H., Schindler, S., \& Yoshikawa, K. 1999;ApJ;516;L1;
\apjpap Lacey, C., \& Cole, S. 1993;MNRAS;262;627;
\apjpap Makino, N., \& Suto, Y. 1993;ApJ;405;1;
\apjpap Markevitch, M., Blumenthal, G. R., Forman, W., Jones,
C., \& Sunyaev, R. A. 1991;ApJ;378;L33;
\apjpap Masai, K. 1984;Ap\&SS;98;367;
\apjpap Mo, H. J., \& White, D. M. 1996;MNRAS;282;347;
\apjpap Mushotzky, R. F., \& Scharf, C. A. 1997;ApJ;482;L13;
\apjpap Nakamura, T. T., \& Suto, Y. 1997;Prog. Theor. Phys.;97;49;
\apjpap Ostriker, J. P., \& Vishniac, E. T. 1986;ApJ;306;L51;
\apjbook Peebles, P. J. E. 1980;The Large Scale Structure of the
Universe;Princeton;Princeton Univ. Press;
\apjpap Persi, F. M., Spergel, D. N., Cen, R., \& Ostriker,
J. P. 1995;ApJ;442;1;
\apjpap Press, W. H., \& Schechter, P. 1974;ApJ;187;425;
\apjpap Readhead, A. C. S., Lawrence, C. R., Myers, S. T., Sargent,
W. L. W., Hardebeck, H. E., \& Moffet, A. T. 1989;ApJ;346;566;
\apjpap Rees, M. J., \& Sciama, D. W. 1968;Nature;517;611;
\apjpap Rosati, P., Della Ceca, R., Norman, C., \& Giacconi, R.
1998;ApJ;492;L21;
\apjpap Shaeffer, R., \& Silk, J. 1988;ApJ;333;509;
\apjpap Subrahmanyan, R., Kesteven, M. J., Ekers, R. D., Sinclair,
M., \& Silk, J. 1998;MNRAS;298;1189;
\apjpap Sugiyama, N. 1995;ApJS;100;281;
\apjpap Sunyaev, R. A., \& Zel'dovich, Ya. B. 1972;
Comments Astrophys. Space Phys.;4;173;
\apjpap White, S. D. M., Navarro, J. F., Evrard, A. E., \& Frenk,
C. S. 1993;Nature;366;429;
\apjpap Zel'dovich, Ya. B., \& Sunyaev, R. A. 1969;Ap\&SS;4;301;

\clearpage

\clearpage

\clearpage

\clearpage

\end{document}